\documentclass[conference]{IEEEtran}
\IEEEoverridecommandlockouts
\usepackage{cite}
\usepackage{amsmath,amssymb,amsfonts}
\usepackage{algorithmic}
\usepackage{graphicx}
\usepackage{textcomp}
\usepackage{xcolor}

\usepackage{multirow}
\usepackage{tabularx}
\usepackage{adjustbox}
\usepackage{ragged2e}

\usepackage{subcaption}
\graphicspath{ {images/} } 

\def\BibTeX{{\rm B\kern-.05em{\sc i\kern-.025em b}\kern-.08em
    T\kern-.1667em\lower.7ex\hbox{E}\kern-.125emX}}
    
\DeclareCaptionFont{mysize}{\fontsize{8}{9.6}\selectfont}
\begin{document}

\title{Automated Empathy Detection for Oncology Encounters}

\author{\IEEEauthorblockN{Zhuohao Chen$^1$, James Gibson$^1$, Ming-Chang Chiu$^1$, Qiaohong Hu$^2$, Tara K. Knight$^3$, \\ Daniella Meeker$^2$, James A. Tulsky$^4$, Kathryn I. Pollak$^5$, Shrikanth Narayanan$^1$}
\IEEEauthorblockA{$^1$Signal Analysis and Interpretation Lab, University of Southern California, Los Angeles, CA, USA \\
$^2$Keck School of Medicine, University of Southern California, Los Angeles, CA, USA\\
$^3$Schaeffer Center for Health Policy \& Economics, University of Southern California, Los Angeles, CA, USA\\
$^4$Dana Farber Cancer Institute and Brigham and Women's Hospital, Harvard Medical School, Boston, MA, USA\\
$^5$Duke Cancer Institute, Durham, NC, USA\\
Email: $^1$sail.usc.edu, $^2$dmeeker@usc.edu, $^3$knight@healthpolicy.usc.edu, $^4$JamesA\_Tulsky@dfci.harvard.edu, \\ $^5$kathryn.pollak@duke.edu}}
\maketitle

\begin{abstract}
Empathy involves understanding other people's situation, perspective, and feelings. In clinical interactions, it helps clinicians establish rapport with a patient and support patient-centered care and decision making. 
Understanding physician communication through observation of audio-recorded encounters is largely carried out with manual annotation and analysis. However, manual annotation has a prohibitively high cost. In this paper, a multimodal system is proposed for the first time to automatically detect empathic interactions in recordings of real-world face-to-face oncology encounters that might accelerate  manual processes.
An automatic speech and language processing pipeline is employed to segment and diarize the audio as well as for transcription of speech into text. Lexical and acoustic features are derived to help detect both empathic opportunities offered by the patient, and the expressed empathy by the oncologist. We make the empathy predictions using Support Vector Machines (SVMs) and evaluate the performance on different combinations of features in terms of average precision (AP).
\end{abstract}


\section{Introduction}

In 2019, over 1.7 million Americans will be diagnosed with cancer, and it will be the cause of death for over half a million people \cite{CancerFacts}. Cancer diagnosis and treatment carries both a physical burden and, for many patients, severe psychological distress \cite{zabora2001prevalence, derogatis1983prevalence}. Conversations between patients and oncologists acknowledge that emotion may reduce this distress \cite{neumann2011identifying, roberts1994influence, takayama2001relationship, spencer2010anxiety, rutter1996doctor, zachariae2003association}. Consensus exists that eliciting patient goals and responding to emotion are important in clinical practice \cite{clayton2013evaluation, back2007efficacy, butcher2016dana}, and patients whose physicians participate in communication training programs may benefit\cite{butcher2016dana, lorenz2008evidence, epstein2005patient}.

A comprehensive systematic review of 39 oncology studies concluded that empathy is associated with higher patient satisfaction, better psychosocial adjustment, lessened psychological distress and the need for information\cite{lelorain2012systematic, morse2008missed}. Many definitions and metrics for empathy and closely related concepts in medical communication have been developed and deliberated. Metrics rely on provider reports, patient reports, and direct observation of conversations using structured coding systems \cite{wilt1998empathy, kunyk2001clarification, rohani2018clinical}. There has been extensive development of coding systems for characterizing patient-provider communications. Training programs that measure and provide feedback about empathic communication are effective in improving communications and outcomes\cite{riess2012empathy, tulsky2011enhancing, clayton2013evaluation}. Effective communication is a learnable clinical skill that can have meaningful consequences on patients' quality of life and decision-making.

A novel approach to training, SCOPE (Studying Communication in Oncologist Patient Encounters), provides personalized feedback to physicians based on manual coding on recorded real-world conversations between patients and providers\cite{koropchak2006studying}. Coded conversations are packaged so that clinicians can review feedback in online performance reports. In a multisite randomized controlled trial, physicians receiving personalized feedback were twice as likely to use empathic statements and to respond appropriately to empathic opportunities\cite{tulsky2011enhancing}, and their patients reported greater trust in their physicians using the 5-point Trust in Physicians Scale\cite{anderson1990development}. While this approach to improving communication with personalized feedback increases empathic responses and patient trust at a fraction of the expense of conventional training, it has limited scalability. Efficiently scaling personalized training while addressing concerns of bias has potential to transform cancer care by improving communication and shared decision-making\cite{pham2014closing}.

This interdisciplinary project is devoted to accelerating the feedback process with automation. Computational methods have promising results that are well-aligned with theories grounded in cognitive science and neuroscience. Work by the SAIL team \cite{xiao2015rate, xiao2016computational, can2016sounds, gibson2015predicting} have focused on a wide range of provider-patient interactions in diagnostic and intervention settings, including psychotherapy\cite{xiao2015analyzing}. Supervised learning algorithms have successfully been used to predict empathy categorizations based on the Motivational Interviewing Skill Code\cite{miller2003manual} and the Motivational Interviewing Treatment Integrity coding systems\cite{moyers2010revised, xiao2012analyzing}. Automated signal processing and machine learning tools that extract features from dialogues associated with empathy\cite{jaffe1970rhythms}, were associated with manually assigned empathy scores\cite{xiao2013modeling, lee2014computing}. Although the basic science is in its early stage, computational empathy analysis has reinforced longstanding theories in cognitive science and neuroscience. For example, empathic interactions have been recognized to have features of entrainment--a synchrony behavior associated with neurobiology as deeply rooted as mirror neurons\cite{gallese2001shared}. 

However, annotation for empathic interactions is still a manual process which has a prohibitively high cost. The coders have to listen to the whole recording carefully to find when the empathic opportunities and responses occur before coding. Building a automatic system to detect the empathic interactions can help reduce the cost of this process. Recently, an empathic conversational system for human-human spoken dialogues was proposed in\cite{alam2018annotating}, that perceives and annotates empathy in customer-agent conversations from the call center corpus in Italian. However, these dialogues involves only one patient and one therapist who speak in separated audio channels. In this paper, we aim to detect empathic interactions in real-world face-to-face oncology conversational recordings, that included multiple speakers mixed in the same channel. We evaluated a multimodal system with a speech and language processing pipeline to 1) segment, diarize and transcribe the audio signals, and 2) extract different types of lexical and acoustic features to predict the empathic interactions. Our result shows the system can filter a subset of the recording which includes most of the empathic interactions.

\section{Data}

\subsection{Corpus}

We use the data from the COPE study\cite{koropchak2006studying, tulsky2011enhancing} (a follow-up study to the original SCOPE trial) in which the audio of 435 oncology encounters, about 164.8 hours in total, were recorded at a 16kHz sampling rate. Among these recordings, 52 conversations are transcribed. All sessions included a patient (PAT), and one or more healthcare providers (HCPs). In some sessions a third party was present, typically friends or family members (FFs) of the patient. Besides audio recordings, the number of total speakers for each session was also provided. A single session had 3.66 speakers on average.

\subsection{Annotation Information}

These oncology sessions were coded by two trained research staff listening to the recordings following the measures described in\cite{pollak2007oncologist}.  Whenever the coder perceives the patient expressing a negative emotion, he records the start time and end time of this empathic interaction which includes both the patient's empathic opportunity and the extent of oncologists' empathic response. 
There are 270 empathic interactions in all 435 sessions. Among these 60 sessions are annotated by both coders, each coder listens to 30 encounters first coded by the other, and then decides whether he agrees with the recorded interactions and points out the empathic interactions not recorded. The coding agreement between coders is kappa=0.71.

\section{Computational System Description}

\begin{figure}[h]
  \centering
  \includegraphics[width=\linewidth]{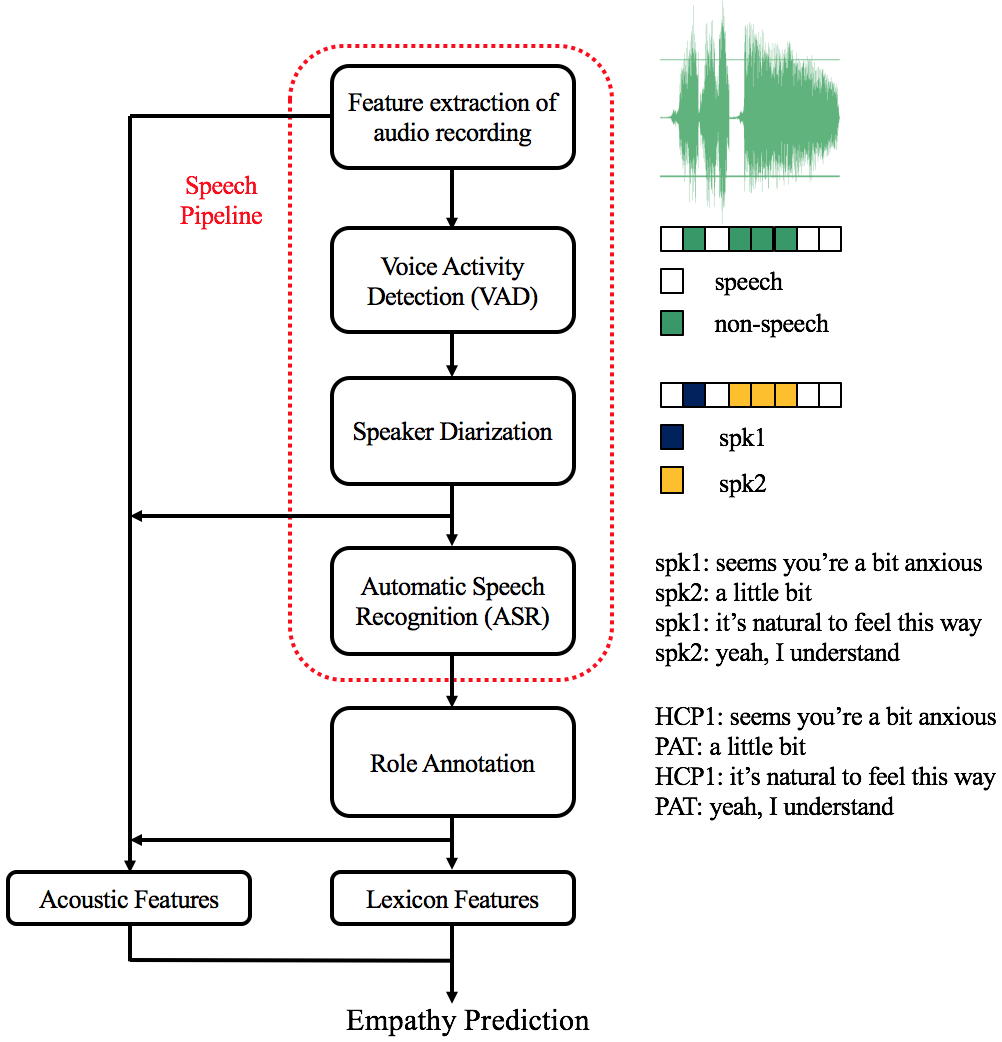}
  \caption{The Multimodal System}
  \label{fig:muitlmodal}
\end{figure}

In the COPE coding process, coders were instructed to identify empathic opportunities based on patient expressions of emotion, both direct and indirect Physicians' responses were categorized primarily based on lexical content.  The empathic interactions were coded based on both lexical and acoustic characteristics. In Fig.~\ref{fig:muitlmodal}, we constructed a multimodal system for empathy prediction. We extracted lexical and acoustic features with the help of a speech pipeline (shown inside the dashed border in the figure). Empathy prediction was made using the acoustic and lexical information. 

\subsection{Speech Processing Pipeline}
The speech pipeline includes a number of components, further described below.

\subsubsection{Voice Activity Detection}

The Voice Activity Detection (VAD) model separates the audio into speech and non-speech part using MFCC features. We pre-trained a 2-layer feed-forward neural network using the DARPA RATs data of noisy speech \cite{walker2012rats}. 

\subsubsection{Speaker Diarization}

The speaker diarization module, followed by VAD, determined who is speaking when in an audio stream. It is challenging due to overlapping speech, rapid speaker changes and non-stationary noise. In this pipeline, we directly used the pre-trained Kaldi model available online\cite{kaldiM6} which is an x-vector diarization model with Probabilistic Linear Discriminant Analysis (PLDA) framework\cite{snyder2017deep}. The Fig.~\ref{fig:muitlmodal} shows that the diarization module clusters speech segments for different speakers, but without role information.

\subsubsection{Automatic Speech Recognition}

The Automatic Speech Recognition (ASR) module transcribes the audio signal features into text. We trained the ASR model with a large combined dataset including Fisher\cite{cieri2004fisher}, Librispeech\cite{panayotov2015librispeech}, Tedlium\cite{rousseau2012ted}, ICSI\cite{janin2003icsi}, AMI\cite{carletta2005ami}, WSJ\cite{doddington1992csr} and  Hub4\cite{graff19971996} using the ASpIRE recipe\cite{peddinti2015jhu}. The language model (LM) is a 3-gram model trained with the SRILM toolkit\cite{stolcke2002srilm}. The model we applied is based on interpolating the LMs trained on Fisher (a corpus of telephone conversations) and a general psychotherapy corpus\cite{press2009counseling} respectively. The mixing weight for Fisher is 0.2. To assess how the speech pipeline performed, we applied the gentle forced aligner\cite{gentle} on the 52 oncology encounter recordings with transcripts to achieve time-stamps for words, and then filled them into the segments to achieve the reference text. The Word Error Rate (WER) of the ASR model was 45.37\%. One of the major sources for the error is that there were always multiple speakers in one session, making the diarization tougher. Nonetheless, the decoded transcripts still retained lexical information useful for empathy detection.

\subsection{Role Annotation}

An empathic interaction contains empathic opportunities and responses between the PAT and HCP. To discern opportunity and response, it is important to know whether an utterance is made by a PAT or a HCP. We collected the number of words for PAT, HCP and FF by the transcribed sessions of 200K words, the ratio of their utterances is 41\%:54\%:5\%. 
The general idea behind role annotation is to train 3-gram LMs of different roles, and for each speaker we find the LM which minimizes the perplexity of his corpus. The utterances of FFs are sparse which are always confused with the ones of PATs, so we attributed FF utterances to PAT. We trained 3-gram background LMs $L_{Pb}$, $L_{Hb}$ for PAT and HCP with the utterances of the therapist and patient of previously mentioned general psychotherapy corpus respectively. 
The 3-gram in-domain LM $L_{Pi}$ of PAT is trained using the PAT and FF's corpus of the transcribed COPE encounters, while the in-domain LM $L_{Hi}$ of HCP is trained by the HCP's corpus of those transcripts. 
The LMs $\widetilde{L}_{P}$ and $\widetilde{L}_{H}$ for PAT and HCP are expressed as

\begin{equation}
L_{P} = \lambda_{1}L_{Pb} \oplus (1-\lambda_{1})L_{Pi}
\end{equation}

\begin{equation}
L_{H} = \lambda_{1}L_{Hb} \oplus (1-\lambda_{1})L_{Hi}
\end{equation}

\begin{equation}
\widetilde{L}_{P} = \lambda_{2}L_{P} \oplus (1-\lambda_{2})L_{H}
\end{equation}

\begin{equation}
\widetilde{L}_{H} = (1-\lambda_{2})L_{P} \oplus \lambda_{2}L_{H}
\end{equation}

where (1) and (2) interpolate the in-domain and background LMs with $\lambda_{1}=0.5$, while the formulas (3) and (4) ensure the two LMs are used with the same vocabulary with $\lambda_{2}=0.01$. 

The 5-fold cross validation annotation result of the 52 transcribed sessions are shown in Table \ref{tab:role}. We find most HCPs and PATs are correctly assigned. For FFs, the majority class is PAT which is consistent with our proposition, though 39\% of them are identified as HCP.

\begin{table}[htbp]
\caption{Role Annotation of Transcribed Sessions}
\begin{center}
  \label{tab:role}
\begin{tabular}{|c|c|c|}
\hline
\multirow{2}{*}{True Role} & \multicolumn{2}{c|}{Predicted Role} \\ \cline{2-3}
  & HCP  & PAT \\ \hline
  HCP   & 52   & 0  \\ \hline
  PAT    & 1    & 83 \\ \hline
  FF   & 13    & 20 \\ \hline
\end{tabular}
\end{center}
\end{table}

\section{Empathy Detection}

\begin{figure}[htb]
  \centering
  \includegraphics[width=9cm]{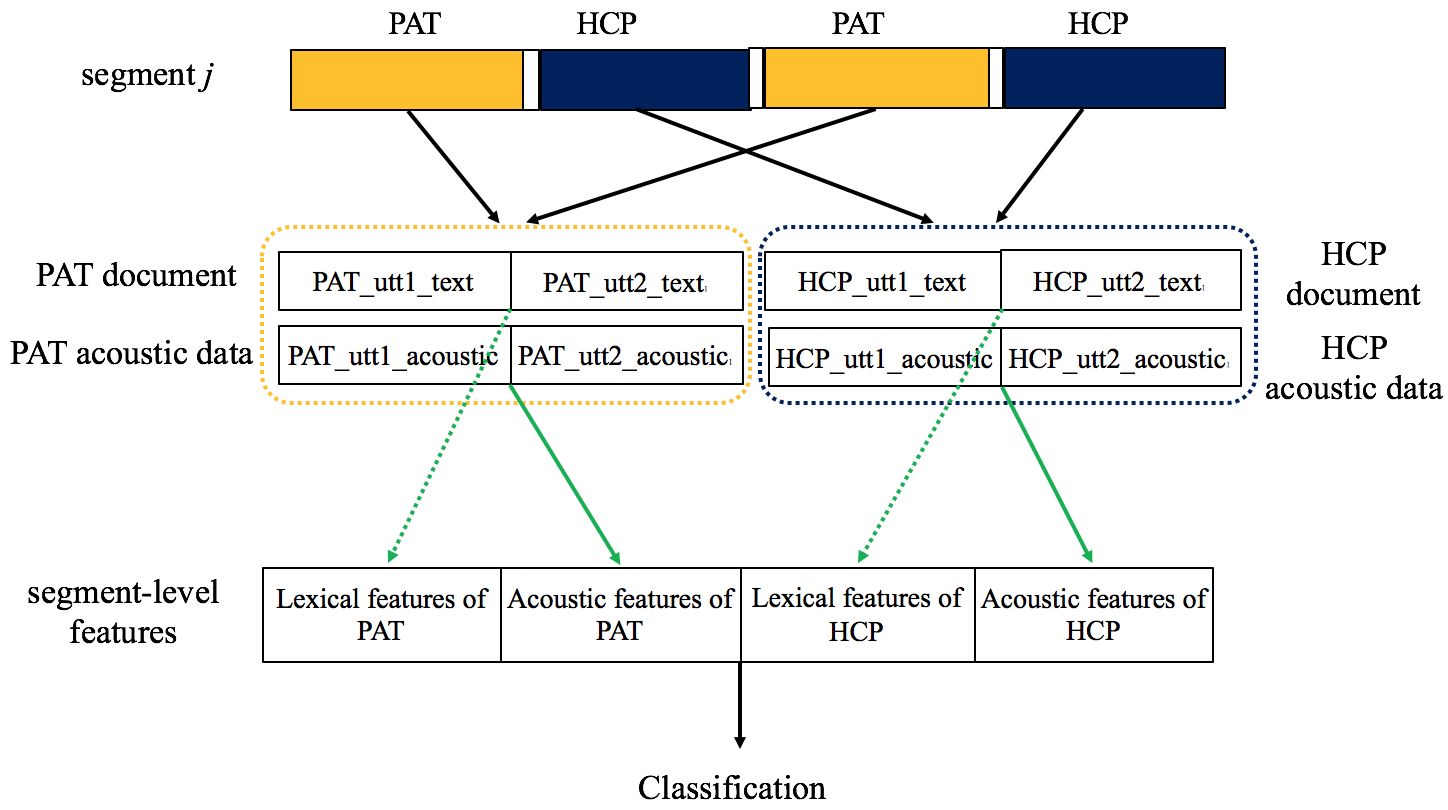}
  \caption{Structure of the Detection Schema}
  \label{fig:classify}
\end{figure}

One of the challenging problems in comparison to prior work is that the manual coding process assigned time stamps for potentially empathic interactions, and the duration of empathic interactions varies from 3 seconds to 93 seconds which is hard to keep track of. So the first step of empathy detection is generating proper training and testing sample segments. Next we extract lexical and acoustic features for both roles in each segment and made multimodal prediction to find empathic interactions by different combinations of features. If there is only one role in one segment, we set a zero vector for the absent role. The structure of the detection schema is shown in Fig.~\ref{fig:classify}.

\subsection{Generating and Labeling Samples}

\begin{figure}[htb]
  \centering
  \includegraphics[width=5cm]{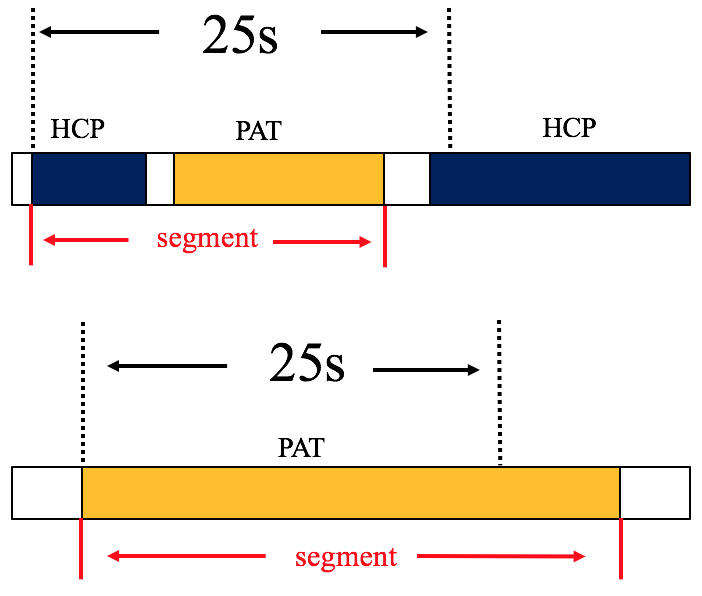}
  \caption{Segment Generating}
  \label{fig:segment}
\end{figure}

The speech pipeline discussed in the previous section provides us the decoded utterances with time stamps and role annotations. Because decoded utterances are usually fragmented, we group the decoded utterances into segments of approximately 25 seconds, which is the average duration of empathic interactions. As shown in Fig.~\ref{fig:segment}, we group the neighboring utterances into a single segment until the the duration of the segment is closest to 25 seconds. For each session, the first group of utterances starts from the beginning of the recording. These extracted segments are samples in our task.

\begin{figure}[htb]
  \centering
  \includegraphics[width=5cm]{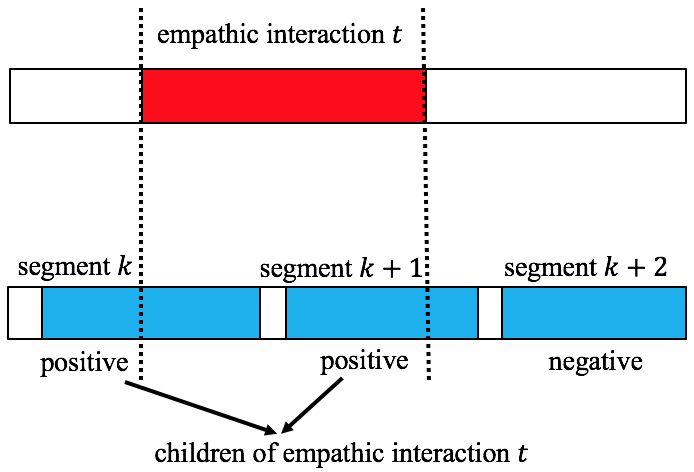}
  \caption{Labeling Segments}
  \label{fig:labeling}
\end{figure}

Having segmented samples generated, we label them as positive or negative according to its overlap with empathic interactions. The Fig.~\ref{fig:labeling} denotes how we assign the labels to segments. We label a segment as positive if it has more than 1 second overlapping time with any empathic interaction. The example in the figure shows that sometimes an empathic interaction can produce more than one positive sample. We call these positive samples the "children" of this empathic interaction. Using this labeling schema we collect 470 positive samples and 21871 negative samples.  From these segments we generate features based on lexical, linguistic, and acoustic properties.

\subsection{Lexical Features}

As presented in Fig.~\ref{fig:classify}, we collected the decoded transcripts for PAT and HCP in each segment. To exploit the lexical characteristics of segments, three types of features are extracted. 

\subsubsection{Doc2vec}

 A sentence embedding approach, doc2vec, learning to represent variable-length pieces of texts with the fixed-length features is adopted \cite{le2014distributed}. It clusters sentences with similar meanings which can be used to learn the general behavior and context represented in the text transcripts. In this paper, we pre-trained the doc2vec with the general psychotherapy corpus and MI corpus and output 100-dimension embedding for each document.  

\subsubsection{Linguistic Inquiry Word Count}

The Linguistic Inquiry Word Count (LIWC)\cite{pennebaker2015development} is a dictionary based text analysis tool which has been widely used to classify texts along linguistic and psychological dimensions and to predict human behaviors. In this paper, we extracted 66 LIWC features for each role of the segment. They include 2 general descriptor categories (total word count, percentage of words captured by the dictionary), 22 standard linguistic dimensions, 32 word categories tapping psychological constructs, 7 personal concern categories and 3 paralinguistic dimensions.

\subsubsection{Empath}

Empath\cite{fast2016empath} is a tool similar to LIWC with much larger lexicon mined on the modern text on web. We took all the pre-built concepts provided by Empath to obtain 194-dimension feature vectors for PAT and HCP documents in each segment.

\subsection{Acoustic Features}

We separated acoustic features into two categories, cepstra (representing segmental speech properties) and prosody (for capturing suprasegmental speech properties). The first group consists of 12 mel-frequency cepstral coefficients (MFCC 1--MFCC 12) which are used in the speech pipeline. The second includes pitch, energy, jitter and shimmer which are relevant to the prosody. The energy feature is presented by MFCC0. Both MFCCs and pitch are extracted using Kaldi toolkit\cite{povey2011kaldi, ghahremani2014pitch}. Jitter and Shimmer are computed by openSMILE\cite{eyben2010opensmile} toolkit. For all the features the frame size is set to 25ms at a shift of 10ms and we applied z-normalize to them in the speaker level. For each role, we concatenate the normalized features and extract their descriptive statistics of max, min, mean, median, standard deviation, skewness and kurtosis.

\begin{table}[htb]
\caption{Features Description}
\begin{center}
 \label{tab:feature}
\begin{tabular}{|c|c|c|}
\hline
\textbf{Feature} & Description                                                                                                                                             & \begin{tabular}[c]{@{}c@{}}Dims for\\  Each Role\end{tabular} \\ \hline
Doc2vec          & document embedding                                                                                                                                      & 100                                                                 \\ \hline
LIWC             & features of psychological states                                                                                                                        & 66                                                                  \\ \hline
Empath           & \begin{tabular}[c]{@{}c@{}}pre-built features generate\\ from common topics \\ on web\end{tabular}                                                      & 194                                                                 \\ \hline
cepstrum          & \begin{tabular}[c]{@{}c@{}}max, min, mean, median, standard\\ deviation, skewness and \\ kurtosis of MFCC1-12\end{tabular}                              & 84                                                                  \\ \hline
Prosodic         & \begin{tabular}[c]{@{}c@{}}max, min, mean, median, standard \\ deviation,skewness and kurtosis of \\ log-pitch, energy, jitter and shimmer\end{tabular} & 28                                                                  \\ \hline
\end{tabular}
\end{center}
\end{table}

\subsection{Prediction Scheme}

The summary of features is shown in Table \ref{tab:feature}. We concatenated different combinations of doc2vec, LIWC, empath, cepstral and prosodic features for both PAT and HCP. Then we trained the the classification model using Support Vector Machines (SVMs). In testing we returned the posterior probabilities \cite{platt1999probabilistic} and sorted out the instances in terms of their probabilities of being a positive empathy sample. 

We randomly split the data into training and testing sets by sessions with ratio of roughly 3:1, while excluding the transcribed sessions from the testing one and making sure no speaker exists in both sets. The data is shown in Table \ref{tab:data}

\begin{table}[htb]
\caption{Frequency of Samples}
\begin{center}
 \label{tab:data}
\begin{tabular}{|c|c|c|c|c|}
  \hline
\textbf{Dataset} & Positive & Negative & Total Samples  & \begin{tabular}[c]{@{}c@{}}empathic\\interactions\end{tabular} \\ \hline
\textbf{Train}   & 341  & 16247 & 16588  & 194 \\ \hline
\textbf{Test}    & 129  & 5624  & 5753   & 76  \\ \hline
\end{tabular}
\end{center}
\end{table}

\section{Experiment}

\subsection{Evaluation Metrics}

The goal of our experiment is to filter out irrelevant content (negative samples) from the recordings while preserving as many empathic interactions as possible. For the evaluation metric, we adopted the average precision (AP), the area under the precision-recall curve, which is claimed to be a better measure of success of prediction than receiver operating characteristic (ROC) curve when the classes are imbalanced\cite{saito2015precision}.

From Section 3.1 we know that some of the positive samples might come from a single empathic interaction. Considering this, we defined the Empathy Detection Rate (EDR) - the \% of the samples needed with the highest probability being positive to recall a certain ratio of empathy interactions. For example, given the output of a model, EDR 50\% at 5\% means that 50\% of the empathy interactions are recalled by 5\% of samples with the highest probability being a positive sample. An empathy interaction is detected if any one of its children is among these selected segments. We report EDR of recalling 20\%, 50\%, and 80\% empathic interactions.

\subsection{Training Routine}

To reduce the class imbalance, we under-sampled the negative instances by 5 in the training set resulting in the ratio between negative:positive being 9.5:1. The SVM module we applied was imported from the scikit-learn package\cite{pedregosa2011scikit}. The sklearn.svm.SVC is implemented with Gaussian kernel and we set the parameter probability=True to enable probability estimates. We tried different configurations of the penalty parameter $C \in [10^{-2}, 10^{-1}, 1]$, kernel coefficient $\gamma \in [10^{-4}, 10^{-3}, 10^{-2}]$ and the class weight negative:positive = $1:W$, where $W \in [1,2,...,10]$. The optimal parameters were selected by 5-fold cross validation.

\subsection{Results and Analysis}

The performances of different feature combinations in terms of average precision is shown in Table \ref{tab:AP}. We set the baseline performance by random prediction, which is equal to the proportion of the positive samples. By comparing Prosody VS Prosodic + Cepstrum, and Doc2vec + LIWC + Empath + Prosody VS Doc2vec + LIWC + Empath + Prosody + Cepstrum we conclude the cepstral features do not help for empathy prediction. We also find the lexical features have a much better performance than acoustic features, while AP of Empath features is the highest among the single-type lexical features. The best result is achieved by combining Doc2vec, LIWC, Empath and Prosody, which obtains the AP of 7.61\% and is much better than the random baseline. The result of AP is generally low because of the sparsity of the empathic interactions in oncology sessions. 

\begin{figure*}[htb]
  \centering
  \includegraphics[width=\textwidth]{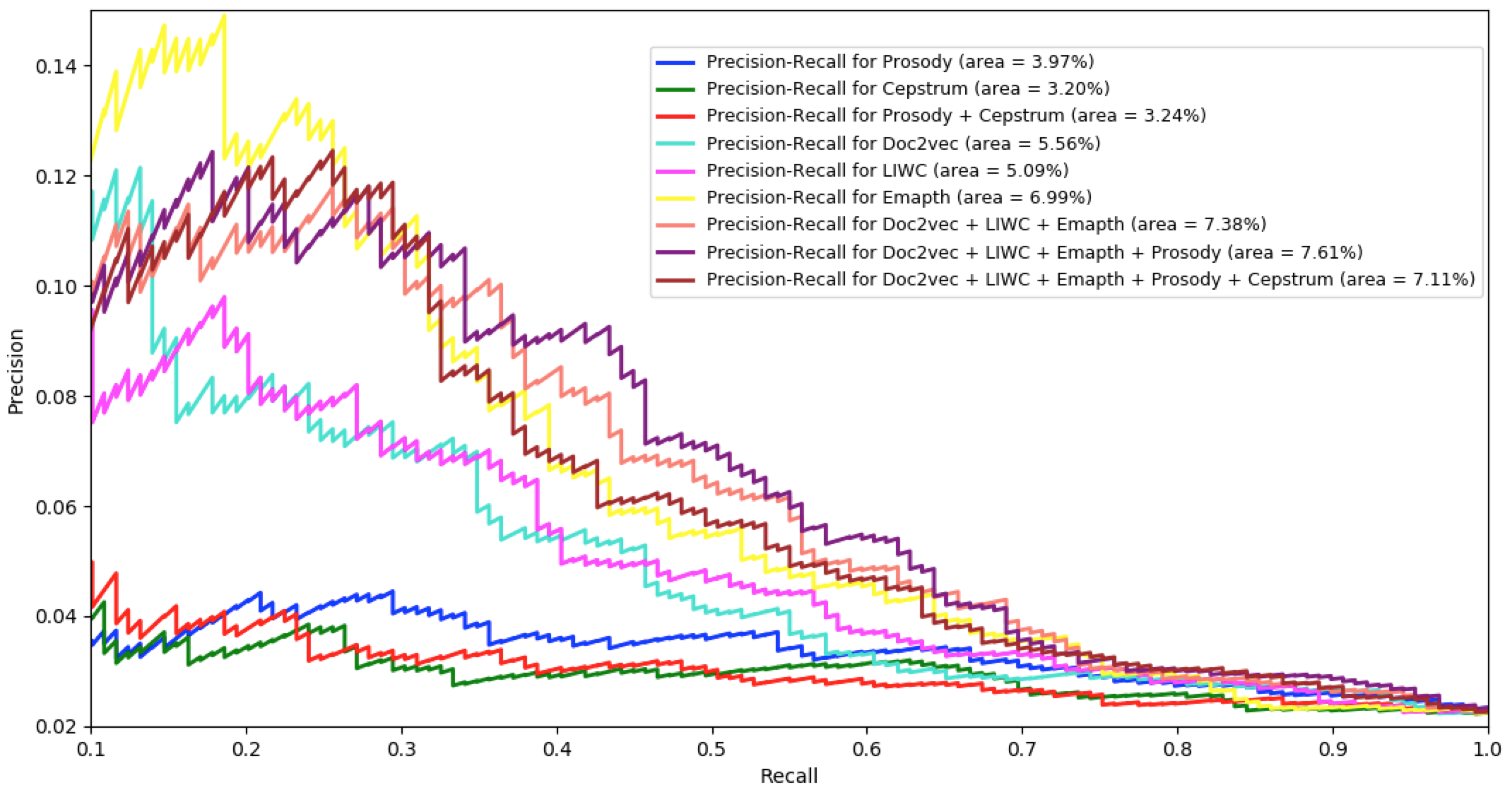}
  \caption{Precision-Recall Curves for Different Feature Combinations}
  \label{fig:curve}
\end{figure*}

A more detailed result is shown in Fig.~\ref{fig:curve}. From this we find that the feature set consisting of Doc2vec, LIWC, Empath and Prosody outperform all the other combinations at most recall levels. When recall $<$ 0.3, the single feature-type prediction of Empath could detect the positive samples more efficiently. The combined lexical features of Doc2vec + LIWC + Empath have comparable performance compared to Doc2vec + LIWC + Empath + Prosody at low recall levels, but degrade quickly as the recall increases. One possible reason is that the positive segments with empathic lexical information are easier to detect, while some of the positive samples, which do not have distinct words conveying PAT's empathic opportunities or HCP's empathy towards PAT, are more likely to be captured by prosodic features.

\begin{table}[htb]
\caption{Average Precision for Different Combined Features}
\begin{center}
 \label{tab:AP}
\begin{tabular}{|c|c|}
\hline
\textbf{Experiment (Feature Combination)}    & \textbf{AP (\%)} \\ \hline
Random Baseline      & 2.10         \\ \hline
Prosody           & 3.97          \\ \hline
Cepstrum             & 3.20           \\ \hline
Prosody + Cepstrum   & 3.24             \\ \hline
Doc2vec             & 5.56             \\ \hline
LIWC                & 5.09             \\ \hline
Empath              & 6.99             \\ \hline
Doc2vec + LIWC + Empath     & 7.38             \\ \hline
Doc2vec + LIWC + Empath + Prosody  & \textbf{7.61}             \\ \hline
Doc2vec + LIWC + Empath + Prosody + Cepstrum & 7.11  \\ \hline
\end{tabular}
\end{center}
\end{table}

The performances of Empathy Detection Rate are shown in Table \ref{tab:rate}. Besides the percentage of samples (POS) needed to recall empathic interactions at different levels, we also present the percentage of length of audio (POA) by summing up the length of the selected samples, and divide it by the total duration of all the audio sessions. The results echo the precision-recall curves in Fig.~\ref{fig:curve}. The outcome of the multimodal prediction from Doc2vec + LIWC + Empath + Prosody shows us that we can detect half of the empathic interactions by listening to only 6.61\% of the recording, and recall 80\% empathic interactions from 23.48\% of the recordings.

\begin{table}[htb]
\centering
\caption{Results of Empathy Detection Rate}
\begin{center}
 \label{tab:rate}
\resizebox{0.5\textwidth}{!}{
\begin{tabular}{|c|c|c|c|c|c|c|}
\hline
Level of Recall                                                                         & \multicolumn{2}{c|}{20\%} & \multicolumn{2}{c|}{50\%} & \multicolumn{2}{c|}{80\%} \\ \hline
Metrics & POS (\%)    & POA (\%)    & POS (\%)     & POA(\%)    & POS (\%)    & POA (\%)    \\ \hline
Prosody  & 9.66  & 9.17 & 23.50 & 22.16 & 48.15 & 45.52 \\ \hline
Cepstrum & 11.98       & 11.36       & 30.73        & 28.75      & 51.80       & 48.47       \\ \hline
Prosody + Cepstrum     & 8.31        & 7.96        & 31.76        & 30.90      & 60.09       & 57.28       \\ \hline
Doc2vec    & 2.43        & 2.33        & 15.19        & 14.71      & 54.49       & 51.25       \\ \hline
LIWC   & 3.53        & 3.31        & 10.41        & 10.01      & 52.62       & 47.28       \\ \hline
Empath  & \textbf{2.07}        & \textbf{1.94}        & 9.49         & 8.90       & 37.96       & 36.10       \\ \hline
Doc2vec + LIWC + Empath  & 2.66        & 2.48        & 7.65         & 7.16       & 30.37       & 28.83       \\ \hline
\begin{tabular}[c]{@{}c@{}}Doc2vec + LIWC + Empath\\  + Prosody\end{tabular}    & 2.62  & 2.42  & \textbf{7.07}  & \textbf{6.61}  & \textbf{24.80}   & \textbf{23.48}       \\ \hline
\begin{tabular}[c]{@{}c@{}}Doc2vec + LIWC + Empath \\ + Prosody + Cepstrum\end{tabular} & 3.06        & 2.83        & 7.51         & 7.04       & 31.03       & 30.04       \\ \hline
\end{tabular}
}
\end{center}
{\justify POS: percentage of samples needed to recall empathic interactions at different levels \\ POA: percentage of length of audio needed to recall empathic interactions at different levels \par}
\end{table}

\section{Conclusion and Future Work}

In this paper, we proposed a system for detecting empathic interactions in real-world oncology conversational recordings. The audio was segmented, diarized and transcribed by a speech processing pipeline from which different types of acoustic and lexical features were extracted. After investigating the effectiveness of acoustic and lexical features and their combinations, we found the combination of Doc2vec (sentence embedding), LIWC (psycholinguistic features), Empath and prosodic features achieved the best performance with a recall rate of 80\% of the empathic interactions from about 23\% of the recording. This result shows that implementing such a multimodal system is a feasible method that might accelerate and ultimately contribute to replacing the manual annotating processes used in SCOPE. It is possible that the effectiveness of SCOPE feedback can be achieved with very few examples; if this is the case, a low-recall approach may be sufficient. The current approach did not attempt to distinguish patient expressions from providers' empathic responsiveness. Future work may further improve upon this by detecting potentially empathic interactions applying lexical and acoustic methods tuned to identify empathic opportunities from patient utterances\cite{sondhi2015vocal, simantiraki2016stress}, which are based on expressions of distress.

\bibliographystyle{IEEEtran}
\bibliography{IEEEexample}

\end{document}